\documentclass[twocolumn,showpacs]{revtex4}

\usepackage[dvips]{graphicx}

\begin{document}

\title{Maximal entanglement of squeezed vacuum states \\
via swapping with number-phase measurement}
\date{\today}
\author{Akira Kitagawa}
\email{kitagawa@nucleng.kyoto-u.ac.jp}
\author{Katsuji Yamamoto}
\email{yamamoto@nucleng.kyoto-u.ac.jp}
\affiliation{Department of Nuclear Engineering,
Kyoto University, Kyoto 606-8501, Japan}

\begin{abstract}
We propose a method to refine entanglement via swapping
from a pair of squeezed vacuum states
by performing the Bell measurement of number sum and phase difference.
The resultant states are maximally entangled
by adjusting the two squeezing parameters to the same value.
We then describe the teleportation of number states
by using the entangled states prepared in this way.
\end{abstract}

\pacs{03.67.-a, 03.67.Hk, 42.50.-p}

\maketitle 

Recently, information technologies with quantum systems have widely
been investigated. Among these advances, quantum teleportation
was first proposed via a protocol by using a two-dimensional system
\cite{BBCJPW}.
This original protocol adopts internal degrees of freedom
as Bell measurement variable, e.g., the polarization of light,
while other protocols were proposed later
to teleport the quadrature phase components
\cite{Braunstein,Milburn}
and the photon number states \cite{Milburn,Cochrane,Cochrane-Milburn}.
Furthermore, the concept of entanglement swapping,
which transfers entanglement from one system to another,
appeared in the verification experiment of teleportation
\cite{Pan}.

In this article, we address the quantum teleportation
in the photon number states.
Great efforts have been done so far theoretically and experimentally
for the teleportation of photon polarization (two dimensional)
and that of quadrature components (infinite dimensional).
Then, it may be desired as an important future step
to realize the teleportation in more than two (but finite) dimensions.
The photon number states seem to be very promising
for such a teleportiation.
The essential ingredients for teleportation
are the EPR (Einstein-Podolsky-Rosen) resource and the Bell measurement.
At present, squeezed vacuum states generated
by parametric down conversion are used as EPR resource
almost uniquely in the experiments utilizing photons,
including the proposals of number state teleportation
\cite{Milburn,Cochrane,Cochrane-Milburn}.
Strictly speaking, however, the squeezed vacuum state is not ideal,
since it has non-uniform number distribution.
Hence, the Fock space of number states should be restricted
effectively to some finite dimensional subspace
in order to prepare physically an ideal EPR resource
for perfect teleportation.
This is suitably achieved by the number sum measurement on two photon modes.
Furthermore, the phase difference is conjugate to the number sum,
as shown later, and they naturally provide the joint Bell measurement.

In this way, it is realized that
the number sum and phase difference measurements
are essential for the multi-dimensional teleportation utilizing photons.
We here formulate properly this Bell measurement
by describing the simultaneous eigenstates
of the Hermitian operators of number sum and phase difference.
Then, we propose a novel method to achieve the desired entanglement
in the number basis.
The Bell measurement plays a crucial role even for this purpose.
Specifically, by performing the number-phase measurement
on the relevant two photon modes,
the EPR states with more favorable distribution
can be obtained via swapping from a pair of squeezed vacuum states.
In particular, by adjusting the two squeezing parameters
to the same value, we obtain the maximally entangled states via swapping
(MESS) with uniform distribution.
These MESS's can be used for a reliable  teleportation protocol
based on the number-phase Bell measurement.

A two-mode squeezed vacuum state is given as
\begin{equation}
| \lambda \rangle_{ab} = ( 1 - \lambda^2 )^{1/2}
\sum_{n=0}^\infty \lambda^n | n \rangle_a | n \rangle_b ,
\label{svs}
\end{equation}
where $ \lambda $ is the squeezing parameter.
This non-uniform distribution with $ \lambda < 1 $ in Eq. (\ref{svs})
is attributed to the fact that the Fock space
is infinite dimensional while the energy should be finite
for the physical states.
In contrast, we will show in the following
that the physical EPR state of maximal entanglement
with uniform distribution may be obtained
if the Fock space is effectively cut off at certain photon number.
We begin with taking the Bell basis
(a complete set of orthonormal entangled states)
for the number sum and phase difference
in the two-mode Fock space $ \{ | n_a \rangle_a | n_b \rangle_b \} $ as:
\begin{equation}
| N , m \rangle_{ab}
= \sum_{k=0}^N \frac{[ ( \omega_{N+1}^* )^m \alpha^* ]^k}{\sqrt{N+1}}
| N - k \rangle_a | k \rangle_b ,
\label{bell-number}
\end{equation}
where $ m = 0 , 1 , \ldots , N $ $ {\rm mod} \ N + 1 $
($ | N , - 1 \rangle_{ab} \equiv | N , N \rangle_{ab} $, etc).
The $ ( N+1 ) $-root to generate
a $ {\rm Z}_{N+1} $ is given by
\begin{equation}
\omega_{N+1} \equiv \exp \left[ i 2 \pi /(N+1) \right] , \
( \omega_{N+1} )^{N+1} = 1 .
\end{equation}
A certain phase factor $ \alpha $ ($ | \alpha | = 1 $) is also introduced,
which will be specified later.
The transformation matrix
$ {\cal U}^{(N)}_{mk}
\equiv [ ( \omega_{N+1}^* )^m \alpha^* ]^k / {\sqrt{N+1}} $
in Eq. (\ref{bell-number}) is really unitary
due to the completeness of the $ ( N+1 ) $-roots
$ 1 , \omega_{N+1} , \ldots , \omega_{N+1}^N $.
These orthonormal Bell states $ | N , m \rangle_{ab} $
are maximally entangled with $ | {\cal U}^{(N)}_{mk} | = 1/{\sqrt{N+1}} $,
spanning the subspace $ \{ | N - k \rangle_a | k \rangle_b \} $
with number sum $ N = n_a + n_b $.
Hence, they serve promisingly as the basis
for $ (N+1) $-dimensional teleportation.

It is here the essential point that these Bell states
are maximally entangled even in terms of the phase states
defined by Pegg and Barnett
\cite{P-B}:
\begin{eqnarray}
| N , m \rangle_{ab}
&=& \exp [ -i \phi_0^{(N,a)} N ]
\nonumber
\\
& \times & \sum_{m^\prime = 0}^N
\frac{[ ( \omega_{N+1}^* )^{m^\prime + m} ]^N}{\sqrt{N+1}}
| \phi^{(N,a)}_{m^\prime + m} \rangle_a
| \phi^{(N,b)}_{m^\prime} \rangle_b .
\nonumber
\\
\label{bell-phase}
\end{eqnarray}
To derive this expression, we take the relations between the number states
and the phase states ($ p = a , b $),
\begin{eqnarray}
&& | n \rangle_p = \sum_{m=0}^N
\frac{\exp [ -i n \phi^{(N,p)}_m ]}{\sqrt{N+1}}
| \phi^{(N,p)}_m \rangle_p ,
\label{number-phase}
\\
&& | \phi^{(N,p)}_m \rangle_p = \sum_{n=0}^N
\frac{\exp [ i n \phi^{(N,p)}_m ]}{\sqrt{N+1}} | n \rangle_p .
\label{phase-number}
\end{eqnarray}
The phases are defined with resolution $ 2 \pi / (N+1) $ as
\begin{equation}
\phi^{(N,p)}_m = \phi^{(N,p)}_0 + \frac{2 \pi}{N+1} m .
\label{phase}
\end{equation}
Here, the superscript $ (N,p) $ denotes the fact
that these phase eigenvalues may include the arbitrary reference phases
$ \phi^{(N,p)}_0 $ possibly depending on the resolutions ($ N $).
By noting $ \exp [ i \phi^{(N,p)}_m ]
= ( \omega_{N+1} )^m \exp [ i \phi^{(N,p)}_0 ] $
in substituting Eq. (\ref{number-phase}) into Eq. (\ref{bell-number}),
the phase factor $ \alpha $ is specified as
\begin{equation}
\alpha = \alpha_{ab} = \exp [ i ( \phi^{(N,a)}_0 - \phi^{(N,b)}_0 ) ]
\end{equation}
for reducing the phase factors
coming from $ \phi^{(N,a)}_0 $ and $ \phi^{(N,b)}_0 $
to the overall $ \exp [ -i \phi_0^{(N,a)} N ] $.
Then, Eq. (\ref{bell-phase}) is led
by using the $ {\rm Z}_{N+1} $ completeness with integer $ m $,
\begin{equation}
\sum_{k=0}^N \frac{[ ( \omega_{N+1} )^m ]^k}{N+1}
= \delta^{(N+1)}_{m 0}
= \left\{ \begin{array}{ll}
1 & ( m = 0 \ {\rm mod} \ N+1 ) \\
0 & ( m \not= 0 \ {\rm mod} \ N+1 ) \end{array} \right. .
\label{cmp-Z}
\end{equation}

The Bell measurement of number sum and phase difference
is represented by the Hermitian operators,
\begin{equation}
{\hat N}_{a+b} \equiv {\hat N}_a + {\hat N}_b , \
{\hat \phi}_{a-b} \equiv \sum_{N=0}^\infty
[ {\hat \phi}^{(N,a)} - {\hat \phi}^{(N,b)} ] {\hat P}_N .
\end{equation}
Here, $ {\hat \phi}^{(N,a)} $ and $ {\hat \phi}^{(N,b)} $
are the phase operators on the modes $ a $ and $ b $, respectively,
providing the eigenvalues of Eq. (\ref{phase}).
The resolution $ 2 \pi / (N+1) $ common to the modes $ a $ and $ b $
should be taken for the phase difference
to be consistent with the number sum.
The projection operator $ {\hat P}_N $ extracts the states
in the subspace $ \{ | N - k \rangle_a | k \rangle_b \} $
with number sum $ N $.
As seen clearly from Eqs. (\ref{bell-number}) and (\ref{bell-phase}),
the Bell states are the simultaneous eigenstates
of number sum and phase difference:
\begin{eqnarray}
&& {\hat N}_{a+b} | N , m \rangle = N | N , m \rangle ,
\label{egn-number}
\\
&& {\hat \phi}_{a-b} | N , m \rangle
= \phi^{(N,a-b)}_m | N , m \rangle ,
\label{egn-phase}
\end{eqnarray}
where the phase difference eigenvalues are given by
\begin{equation}
\phi^{(N,a-b)}_m = [ \phi^{(N,a)}_0 - \phi^{(N,b)}_0 ]
+ \frac{2 \pi}{N+1} m .
\end{equation}
Since $ [ {\hat \phi}^{(N,a)} - {\hat \phi}^{(N,b)} ] $ does not change
the number sum $ N $, it commutes with $ {\hat P}_N $ as required
for the Hermiticity of the entire phase difference operator
$ {\hat \phi}_{a-b} $.
These results clarify that in the subspace with number sum $ N $
the phase difference operator introduced by Luis and S\'anchez-Soto
\cite{L-S} indeed coincides
with the difference of the phase operators of the individual modes
a la Pegg and Barnett \cite{P-B},
while it is not separable in the entire two-mode Fock space.
It is also obvious from Eqs. (\ref{egn-number}) and (\ref{egn-phase})
that $ {\hat N}_{a+b} $ and $ {\hat \phi}_{a-b} $ are commutable:
\begin{equation}
[ {\hat N}_{a+b} , {\hat \phi}_{a-b} ] = 0 .
\end{equation}

\begin{figure}[t]
\scalebox{.4}{\includegraphics*[0cm,7cm][20cm,21cm]{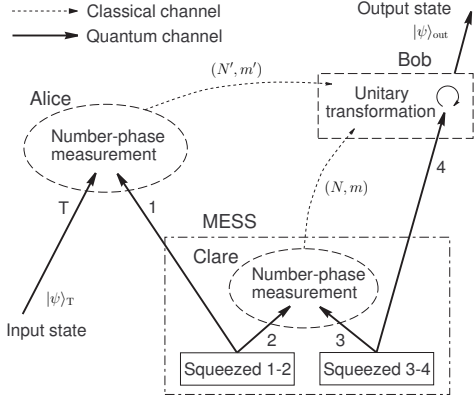}}
\caption{
Quantum teleportation with MESS.
Alice and Bob share MESS, which is prepared by Clare
with the number-phase measurement.
Alice also performs the number-phase measurement.
Bob receives classically the messages from Alice and Clare,
and performs the unitary transformation
to reconstruct the target state.
}
\label{QT-MESS}
\end{figure}

Now we prepare a pair of squeezed vacuum states, 1-2 system and 3-4 system,
for entanglement and teleportation.
They can in fact be expressed in an entangled form
via swapping (1-2, 3-4) $ \rightarrow $ (1-4, 2-3) as
\begin{eqnarray}
| \lambda \rangle_{12} | \lambda^\prime \rangle_{34}
&=& ( 1 - \lambda^2 )^{1/2} ( 1 - \lambda^{\prime 2} )^{1/2}
\nonumber
\\
& \times &
\sum_{N=0}^\infty \sum_{k=0}^N \lambda^{N-k} \lambda^{\prime k}
| N - k \rangle_1 | k \rangle_4 | N - k \rangle_2 | k \rangle_3
\nonumber
\\
&=& ( 1 - \lambda^2 )^{1/2} ( 1 - \lambda^{\prime 2} )^{1/2}
\nonumber
\\
& \times & \sum_{N=0}^\infty \lambda^N
\sum_{m=0}^N | N , m \rangle_{23} | N , - m \rangle_{14}^{(r)}
\label{1234-bell}
\end{eqnarray}
with the generalized Bell states including
the ratio of the squeezing parameters $ r \equiv \lambda^\prime / \lambda $,
\begin{equation}
| N , m \rangle_{14}^{(r)}
= \sum_{k=0}^N r^k
\frac{[ ( \omega_{N+1}^* )^m \alpha_{14}^* ]^k}{\sqrt{N+1}}
| N - k \rangle_1 | k \rangle_4 ,
\end{equation}
where $ \alpha_{14} = \alpha_{23}^* $.
The relation (\ref{1234-bell}) is derived
by extending the sum over $ k $ with $ \delta_{k k^\prime}^{(N+1)} $ as
\begin{equation}
\sum_{k=0}^N ( \cdots ) ( \alpha_{23}^* \alpha_{23} )^k
= \sum_{k=0}^N \sum_{k^\prime = 0}^N
\delta_{k k^\prime}^{(N+1)} ( \cdots )
( \alpha_{23}^* )^k ( \alpha_{23} )^{k^\prime}
\end{equation}
and by taking the completeness of $ {\rm Z}_{N+1} $
for $ \delta_{k k^\prime}^{(N+1)} = \delta_{( k - k^\prime ) 0}^{(N+1)} $
as given in Eq. (\ref{cmp-Z}).
Then, the Bell states, maximally entangled
with $ \lambda = \lambda^\prime $, are obtained in Eq. (\ref{1234-bell})
by performing the number-phase Bell measurement
on the 2-3 system, as shown in Fig. \ref{QT-MESS}
as a part of quantum teleportation with MESS:
\begin{equation}
| \lambda \rangle_{12} | \lambda^\prime \rangle_{34} \Rightarrow
| N , - m \rangle_{14}^{(r)}
\stackrel{\lambda = \lambda^\prime}{\longrightarrow}
| N , - m \rangle_{14} .
\end{equation}
The probability to obtain the result $ ( N , m ) $
with $ \lambda = \lambda^\prime $ in this Bell measurement
is given from Eq. (\ref{1234-bell}) by
\begin{equation}
P ( N , m , \lambda ) = ( 1 - \lambda^2 )^2 \lambda^{2N} .
\end{equation}

The original squeezed vacuum states have the non-uniform distribution
with weight $ \propto ( \lambda^n )^2 $.
In contrast, the resultant Bell states $ | N , - m \rangle_{14} $
with $ \lambda = \lambda^\prime $ are maximally entangled,
having the uniform distribution
in $ ( n_1 , n_4 ) = ( N - k , k ) $ of the modes 1 and 4.
Hence, they provide the ideal EPR resource for the teleportation
of photon number states, as shown in Fig. \ref{QT-MESS}.
The target state may generally be prepared as
\begin{equation}
| \psi \rangle_{\rm T} = \sum_{n=0}^\infty c_n | n \rangle_{\rm T} .
\end{equation}
Alice and Bob share the Bell state
$ | N , - m \rangle_{14}^{(r)} $ of the 1-4 system,
which is prepared by Clare with the number-phase measurement
on the modes 2 and 3 giving the result $ ( N , m ) $.
Then, Alice performs the number-phase measurement
on the modes T and 1 giving the result $ ( N^\prime , m^\prime ) $,
and the output state in the mode 4 is left for Bob as
\begin{eqnarray}
| \psi^\prime \rangle_{\rm out}
&=& {}_{1{\rm T}} \langle N^\prime , m^\prime |
| \psi \rangle_{\rm T} | N , - m \rangle_{14}^{(r)}
\nonumber
\\
&=& \sum_{n=n_0}^{N^\prime}
\frac{\exp [ - i \phi ( N , m , N^\prime,m^\prime ) ( n - \Delta N ) ]}
{\sqrt{N+1}{\sqrt{N^\prime + 1}}}
\nonumber
\\
& \times & [ ( \omega_{N^\prime + 1} )^{m^\prime}
               \alpha_{1{\rm T}} ]^{\Delta N}
r^{n - \Delta N} c_n | n - \Delta N \rangle_4 \
\end{eqnarray}
with $ \Delta N \equiv N^\prime - N $,
$ n_0 \equiv {\rm max}[0,\Delta N] $ and
\begin{equation}
\phi ( N , m , N^\prime,m^\prime ) \equiv
\arg [ ( \omega_{N^\prime + 1}^* )^{m^\prime} \alpha_{1 {\rm T}}^*
( \omega_{N+1}^* )^m \alpha_{14} ] .
\end{equation}
Here, due to the condition $ 0 \leq n - \Delta N \leq N $
for the mode 4, the sum is taken over the photon number $ n $ as
\begin{equation}
n = \left\{ \begin{array}{ll} 0 , \ldots , N^\prime & ( N^\prime \leq N )
\\ \Delta N , \ldots , N^\prime & ( N^\prime > N ) \end{array} \right. .
\label{n-range}
\end{equation}
We have explicitly introduced Clare who prepares the EPR resource,
since in the present method the entangled state is obtained
depending on the result of Bell measurement.
In contrast, it is usually the implicit recognition that
one certain EPR resource, e.g., the singlet state of photon polarizations,
is prepared automatically by some  device,
which is known preceding to a teleportation experiment.
In any case, the information on the EPR resource
should be told to Bob.

Clare informs Bob what Bell state is prepared with the result $ ( N , m ) $,
and Alice tells Bob the result $ ( N^\prime , m^\prime ) $
of Bell measurement.
Then, according to these classical messages
Bob performs a unitary transformation
$ | \psi \rangle_{\rm out} = U_{\Delta N} U_\phi
| \psi^\prime \rangle_{\rm out}$
to reconstruct the original state as faithfully as possible.
The phase $ \phi ( N , m , N^\prime,m^\prime ) $ can be removed
by an operation with number operator as
\begin{equation}
U_\phi = \exp [ i {\hat N}_4 \phi ( N , m , N^\prime,m^\prime ) ] .
\end{equation}
Then, the number shift of $ \Delta N $ can be made
by another operation with phase operator
\cite{P-B} as
\begin{equation}
U_{\Delta N}
= \exp [ - i {\hat \phi}^{( {\rm max} [ N , N^\prime ] , 4 )} \Delta N ] .
\end{equation}
As a result, the output state is obtained as
\begin{equation}
| \psi \rangle_{\rm out}
= \frac{[ ( \omega_{N^\prime + 1} )^{m^\prime}
              \alpha_{1{\rm T}} / r ]^{\Delta N}}
{\sqrt{N+1}{\sqrt{N^\prime + 1}}}
\sum_{n=n_0}^{N^\prime} r^n c_n | n \rangle_4 .
\end{equation}
That is, the number state teleportation is made as
\begin{equation}
c_n \ ( 0 \leq n < \infty ) \ \Rightarrow \
r^n c_n \ ( {\rm max}[ 0 , \Delta N ] \leq n \leq N^\prime )
\end{equation}
up to the physically irrelevant overall phase factor.
In particular, by adjusting $ \lambda = \lambda^\prime $ ($ r = 1 $),
the target state is faithfully reproduced
in the range of photon number 0 to $ N^\prime $
for the case of $ \Delta N \leq 0 $ ($ N^\prime \leq N $).
In other words, if the target state is prepared
in the range 0 to certain maximal number $ {\bar N} $,
the probability for its successful teleportation
with $ {\bar N} \leq N^\prime \leq N $ is given by
\begin{equation}
Q ( {\bar N} , \lambda )
= \sum_{N^\prime={\bar N}}^\infty
\sum_{m^\prime=0}^{N^\prime}
\sum_{N=N^\prime}^\infty
\sum_{m=0}^N
\frac{P ( N , m , \lambda )}{( N + 1 ) ( N^\prime + 1 )}
= \lambda^{2{\bar N}} .
\label{Q}
\end{equation}

We finally discuss possible experimental realizations
of our method to prepare the MESS's
and the number state teleportation with them
($ \lambda = \lambda^\prime $ for definiteness).
The case of $ N = N^\prime = 1 $
may be regarded as the basis-reduced version
of the scheme of Bennett et al
\cite{BBCJPW}.
In fact, we have the two Bell states
$ | N^\prime = 1 , m^\prime = 0 \rangle_{1{\rm T}} $
and $ | N^\prime = 1 , m^\prime = 1 \rangle_{1{\rm T}} $ rather than four.
The number of Bell states is reduced by half
if only the EPR resource with $ N = 1 $ is used
upon the Bell measurement by Clare.
The number-phase measurement can be realized for $ N = 1 $
with beam splitter and phase shift
generating a unitary transformation of the modes 2 and 3 as
\begin{eqnarray}
\left( \begin{array}{c} a_2 ^\dagger \\ a_3 ^\dagger \end{array} \right)
= \left( \begin{array}{cc}
c & -s \eta \\ s \xi & c \eta \xi \end{array} \right)_{(23)}
\left( \begin{array}{c} a_{2^\prime}^\dagger \\
                        a_{3^\prime}^\dagger \end{array} \right) .
\label{U(2)}
\end{eqnarray}
By taking $ c = 1/{\sqrt 2} $, $ s = 1/{\sqrt 2} $,
$ \eta = - 1 $, $ \xi = 1 $,
the Bell states ($ N = 1 , m = 0,1 $) are transformed as
\begin{equation}
| 1 , 0 \rangle_{23} = | 1 \rangle_{2^\prime} | 0 \rangle_{3^\prime} , \
| 1 , 1 \rangle_{23} = | 0 \rangle_{2^\prime} | 1 \rangle_{3^\prime} .
\end{equation}
Then, by making the single photon countings
on the modes $ 2^\prime $ and $ 3^\prime $,
we can identify either of the Bell states with $ N = 1 $,
obtaining the EPR resource $ | 1, 0 \rangle_{14} $
or $ | 1, 1 \rangle_{14} $.
The same technique can be used on the modes T and 1,
and teleportation of a qubit
$ | \psi \rangle_{\rm T}
= c_0 | 0 \rangle_{\rm T} + c_1 | 1 \rangle_{\rm T} $
is completed in the probability
$ P ( 1 , m , \lambda ) = ( 1 - \lambda^2 )^2 \lambda^2 $
(e.g., $ 0.14 $ for $ \lambda = 0.5 $),
as seen from Eq. (\ref{Q}) with fixed $ N = N^\prime = 1 $.
The other results of the photon counters are discarded
for this minimal protocol.
It is here interesting to mention that based on a similar idea
an experimental result has been reported recently
for the teleportation of the vacuum-one-photon qubit
\cite{LSPM}.

The Bell measurements may be done partially even for $ N = 2 $.
The point is to introduce an {\it ancilla} mode $ {\tilde{\rm a}} $.
Then, by making a series of unitary transformations such as
in Eq. (\ref{U(2)}) among the modes 2, 3 and $ {\tilde{\rm a}} $,
it is possible to realize that only
the Bell state $ | N = 2, m = 0 \rangle_{23} | 0 \rangle_{\tilde{\rm a}} $
has the component of $ | 0 \rangle_{2^\prime} | 1 \rangle_{3^\prime}
| 1 \rangle_{{\tilde{\rm a}}^\prime} $,
while the other Bell states
$ | N = 2, m = 1,2 \rangle_{23} | 0 \rangle_{\tilde{\rm a}} $ do not.
This means that if we observe the state
$ | 0 \rangle_{2^\prime} | 1 \rangle_{3^\prime}
| 1 \rangle_{{\tilde{\rm a}}^\prime} $ with single photon detectors,
we can identify at least
$ | N = 2, m = 0 \rangle_{23} | 0 \rangle_{\tilde{\rm a}} $,
producing the EPR resource $ | N=2, m=0 \rangle_{14} $.
The relevant parameters are taken for the sequential transformations as
$ c = 1/{\sqrt 2} , s = 1/{\sqrt 2} , \eta = 1 , \xi = 1 $
for $ (2 {\tilde{\rm a}}) $
$ \rightarrow $
$ c = {\sqrt{2/3}} , s = 1/{\sqrt 3} ,
\eta = 1 , \xi = ( 1 + i )/{\sqrt 2} $
for $ (3 {\tilde{\rm a}}) $
$ \rightarrow $
$ c = {\sqrt{3/8}} , s = - {\sqrt{5/8}} ,
\eta = 1 , \xi = ( 3 + i )/{\sqrt{10}} $
for $ (23) $.
This procedure may be extended for $ N > 2 $.

We believe that the present analysis promotes
the theoretical and experimental efforts
for the number sum and phase difference measurements,
which are essential for realizing the $ (N+1) $-dimensional entanglement
and teleportation utilizing photons.
It is here notable that some ideas and attempts have appeared recently
for the phase difference measurement
\cite{pdm}.
While the partial Bell measurements may be done
with linear operations such as beam splitters and phase shifts,
some non-linear operations will be needed for the perfect Bell measurement,
which is one of very challenging issues in the future
for fundamental quantum physics.

In summary, we have formulated the Bell measurement
in the photon number basis by describing the simultaneous eigenstates
of number sum and phase difference.
Then, we have proposed a novel method to refine entanglement via swapping
from a pair of squeezed vacuum states
by performing this number-phase Bell measurement.
By adjusting the two squeezing parameters to the same value,
these states are maximally entangled.
These MESS's can really be used for the reliable teleportation
of number states
based on the number-phase Bell measurement.
We have also discussed some feasible experimental realizations
of the number-phase measurement for preparing MESS's
and performing teleportation utilizing them,
where beam splitters with phase shifts
and single photon counters may be used.

The authors would like to thank M. Tada for valuable discussions.


\begin{thebibliography}{99}
\bibitem{BBCJPW}
C.~H.~Bennett, G.~Brassard, C.~Crepeau, R.~Jozsa, A.~Peres
and W.~K.~Wootters,
Phys. Rev. Lett. \textbf{70}, 1895 (1993).

\bibitem{Braunstein}
S.~L.~Braunstein and H.~J.~Kimble, Phys. Rev. Lett. \textbf{80}, 869 (1998);
L.~Vaidman, Phys. Rev. A \textbf{49}, 1473 (1994).

\bibitem{Milburn}
G.~J.~Milburn and S.~L.~Braunstein, Phys. Rev. A \textbf{60}, 937 (1999).

\bibitem{Cochrane}
P.~T.~Cochrane, G.~J.~Milburn and W.~J.~Munro,
Phys. Rev. A, \textbf{62}, 062307 (2000).

\bibitem{Cochrane-Milburn}
P.~T.~Cochrane and G.~J.~Milburn,
Phys. Rev. A, \textbf{64}, 062312 (2001).

\bibitem{Pan}
J.-~W.~Pan, D.~Bouwmeester, H.~Weinfurter and A.~Zeilinger,
Phys. Rev. Lett. \textbf{80}, 3891 (1998).

\bibitem{P-B}
D.~T.~Pegg and S.~M.~Barnett, Phys. Rev. A \textbf{39}, 1665 (1989);
S.~M.~Barnett and D.~T.~Pegg, Phys. Rev. A \textbf{42}, 6713 (1990).

\bibitem{L-S}
A. Luis and L.~L. S\'anchez-Soto,
Phys. Rev. A \textbf{48}, 4702 (1993).

\bibitem{LSPM}
E.~Lombardi, F.~Sciarrino, S.~Popescu and F.~De Martini,
Phys. Rev. Lett. \textbf{88}, 070402 (2002).

\bibitem{pdm}
G. Bj\"ork and J. S\"oderholm,
J. Opt. B: Quantum Semiclass. Opt. \textbf{1}, 315 (1999);
A. Trifonov, T. Tsegaye, G. Bj\"ork, J. S\"oderholm,
E. Goober, M. Atat\"ure and A. V. Sergienko,
J. Opt. B: Quantum Semiclass. Opt. \textbf{2}, 105 (2000).

\end{thebibliography}
\end{document}